\documentclass[aps,prc,twocolumn,superscriptaddress,showpacs,floatfix]{revtex4-1}

\usepackage[T1]{fontenc}
\usepackage{latexsym}
\usepackage{graphicx,amssymb}
\usepackage[fleqn]{amsmath}
\usepackage{amsfonts}
\usepackage{dcolumn}
\usepackage{bm}
\usepackage{braket}
\usepackage{multirow}
\usepackage{MnSymbol}
\usepackage{color}
\usepackage{ulem}
\usepackage{hyperref}
\usepackage{bm}

\begin{document}

\preprint{ }

\title{$^{19}$F$(p,\gamma)$$^{20}$Ne reaction rate and the puzzling calcium abundance in metal poor stars }

\author{G.X. Dong}
\affiliation{School of Science, Huzhou University, Huzhou 313000, China}

\author{X.B. Wang}
\affiliation{School of Science, Huzhou University, Huzhou 313000, China}

\author{N. Michel}
\affiliation{CAS Key Laboratory of High Precision Nuclear Spectroscopy,Institute of Modern Physics, Chinese Academy of Sciences, Lanzhou 730000, China}
\affiliation{School of Nuclear Science and Technology, University of Chinese Academy of Sciences, Beijing 100049, China}

\author{M. P{\l}oszajczak}
\thanks{marek.ploszajczak@ganil.fr}
\affiliation{Grand Acc\'el\'erateur National d'Ions Lourds (GANIL), CEA/DSM - CNRS/IN2P3,
BP 55027, F-14076 Caen Cedex, France}

\date{\today}

\begin{abstract}
The $^{19}$F$(p,\gamma)$$^{20}$Ne reaction is the only process to break out of the CNO cycle at temperature below 0.1 GK and may serve as the origin of calcium in first generation of stars after the Big Bang. In the recent measurement, the Jinping Underground Nuclear Experiment (JUNA) obtained the rate of $^{19}$F$(p,\gamma)$$^{20}$Ne reaction, significantly larger than the previously recommended values. In this work, we perform the theoretical studies of the $^{19}$F$(p,\gamma)$$^{20}$Ne reaction using the Gamow shell model in the coupled-channel representation (GSM-CC).
 At temperature around 0.1 GK, the predicted rate by GSM-CC is close to the rate found by JUNA. Thus, based on GSM-CC, the break-out reaction $^{19}$F$(p,\gamma)$$^{20}$Ne from the CNO-cycle might win over its competing reaction $^{19}$F$(p,\alpha)$$^{16}$O, and produce enough calcium in the metal poor stars.
\end{abstract}

\maketitle

\textit{Introduction --}
The CNO cycle is a catalytic fusion process, in which a helium nucleus is formed from the four proton capture process followed by two $\beta^+$ decays, without the loss of CNO seeds. In the CNO cycle, $^{19}$F can be produced through the $^{18}$O$(p,\gamma)$$^{19}$F reaction~\cite{Wirzba,Wiescher80}. The reaction of $^{19}$F$(p,\alpha)$$^{16}$O, would transform $^{19}$F back into the CNO materials. The $^{19}$F$(p,\gamma)$$^{20}$Ne reaction is served as its competitive reaction, which is the only breakout reaction from the CNO cycles at the stellar temperature below 0.1 GK~\cite{Wiescher99}. This energy range is important for the hydrogen burning in main-sequence stars. It could also be important for asymptotic giant branch (AGB) stars, novae, and x-ray bursts~\cite{Wiescher99}. The $^{19}$F$(p,\gamma)$$^{20}$Ne reaction is an irreversible flow from the CNO to the Ne--Na materials, since the backprocess through $^{22}$Ne$(p,\alpha)$$^{19}$F is energetically forbidden. Based on the stellar models, this breakout reaction
may be the source of the calcium abundance in the most metal-poor stars~\cite{Keller14}. However, the rate of $^{19}$F$(p,\gamma)$$^{20}$Ne reaction is believed to be much weaker than the rate of the competing $^{19}$F$(p,\alpha)$$^{16}$O. Using the Nuclear Astrophysics Compilation of Reaction Rates (NACRE)~\cite{nacre}, the calcium abundance would be nearly two orders of magnitude lower than the observations in the most metal-poor stars~\cite{Clarkson21}. Thus, the pinning down of the reaction rates of $^{19}$F$(p,\gamma)$$^{20}$Ne at temperature around 0.1 GK should be very important for the understanding of the calcium abundance and the validation of the stellar models.

Very limited measurements have been done in the low energy region due to the strong background of 6.125 MeV $\gamma$-ray from the $^{19}$F$(p,\alpha)$$^{16}$O reaction, which makes the measurements of the small reaction cross section of $^{19}$F$(p,\gamma)$$^{20}$Ne extremely difficult. Most previous experiments~\cite{Sinclair,Farney,Keszthelyi,Berkes,Subotic} relied on the detection of larger than 11 MeV $\gamma$ transition to the first excited state of $^{20}$Ne.
Later, a coincident detection method was developed to measure $^{19}$F$(p,\gamma)$$^{20}$Ne and $^{19}$F$(p,\alpha)$$^{16}$O reactions over the energy range of center of mass (c.m.) energy $E_{\rm c.m.}$=200--760 keV~\cite{CO08}. Due to the limited sensitivity in the low energy region, only an upper limit was given for the strength of the $2^-$ resonance at $E_{\rm c.m.}$=213 keV, and no estimation was given for the $3^-$ resonance at $E_{\rm c.m.}$=225 keV~\cite{CO08}. More recently, a factor of two larger strength of the $1^+$ resonance at $E_{\rm c.m.}$=323 keV was reported from the measurement in inverse kinematics~\cite{Williams}.
In this measurement, the direct capture to the ground state of $^{20}$Ne has also been reported. Then, deBoer \textit{et al.}~\cite{deBoer} applied the phenomenological $R$-matrix approach to evaluate the previous experimental data sets on $^{19}$F$(p,\gamma)$$^{20}$Ne and $^{19}$F$(p,\alpha)$$^{16}$O reactions, so as to make better characterization of the reaction rate uncertainties. They obtained the ratio of the rates four times lower than that of NACRE. Thus, the problem of the origin of calcium observed in the metal-poor stars became even more puzzling.

Most recently, Jinping Underground Nuclear Experiment (JUNA) Collaboration has done the direct measurement of $^{19}$F$(p,\gamma)$$^{20}$Ne reaction cross section in the underground laboratory in the energy range $E_{\rm c.m.}$=186--292 keV. The $3^-$ resonance at $E_{\rm c.m.}$=225 keV was found to give an important contribution to the reaction rate at around 0.1 GK~\cite{JUNA}. The rate of $^{19}$F$(p,\gamma)$$^{20}$Ne reaction
at $T_9$ = 0.1 reported by JUNA collaboration is  7.4 times larger than that of NACRE, so that the calcium abundance in the first stars could be compatible with the observation.

Most theoretical studies of $^{19}$F$(p,\gamma)$$^{20}$Ne reaction are in the framework of either the $R$-matrix method~\cite{deBoer,JUNA} or the direct capture model~\cite{Wiescher99}. In this Letter, we perform the first microscopic study of this reaction using the Gamow shell model in the coupled-channel representation (GSM-CC)~\cite{GSMbook}, which provides the unified theory of nuclear structure and reactions. It has been used previously for the proton elastic and inelastic scattering~\cite{Jaganathen14}, the deuteron elastic scattering~\cite{Mercenne19}, the triton and $^3$He elastic scattering~\cite{Fernandez23}, and the radiative capture reactions~\cite{Fossez15,dong17,dong22,dong23}.

\textit{Coupled-channel representation of the Gamow shell model --} Here, we briefly introduce the essential steps of the GSM-CC approach. More details can be found in the Supplemental Material~\cite{supplement} (including Refs~\cite{GSMbook,Michel09,Ikeda88,rf:4,Michel02RIB,Michel03,Mercenne19,Furutani78,Furutani79,jaganathen_2017,nndc,Betts75,Kious90,CO08,deBoer,JUNA}).

In GSM-CC, the ${ A }$-body system is described by the reaction channels:%
\begin{equation}
  \ket{ { \Psi }_{ M }^{ J } } = \sum_{ {c} } \int_{ 0 }^{ +\infty } \ket{{ \left( r,c \right) }_{ M }^{ J } } \frac{ { u }_{ {c} }^{JM} (r) }{ r } { r }^{ 2 } ~ dr \; ,
  \label{scat_A_body_compound}
\end{equation}
where ${ {u}_{ {c} }^{JM}(r) }$ is the radial amplitude of the relative motion between the target and projectile in the channel $c$, and ${ r }$ is the relative distance between the c.m. of the target core and the projectile. ${ {u}_{ {c} }^{JM}(r) }$ is obtained by solving the coupled-channel equations for fixed total angular momentum $ {J} $ and projection $ {M} $.
$\ket{ \left( r,c \right)^J_M} $ denotes the binary-cluster channel:
\begin{equation}
  \ket{ \left( r,c \right)^J_M}  = \hat{ \mathcal{A}} \ket{ \{ \ket{ \Psi_{ {\rm T} }^{J_{ {\rm T} } }  }
  \otimes \ket{ {\Psi^{J_{{\rm p}}}_{{\rm p}}}} \}_{ M }^{J} } \; .
  \label{channel}
\end{equation}
The channel index $c$ contains information about binary mass partition and quantum numbers, and the variable $r$ is included in $\ket{ {\Psi_{{\rm p}}^{J_{{\rm p}}}}} $. ${\hat{ \mathcal{A}}}$ stands for the antisymmetrization of nucleons in different clusters. $\ket{\Psi_{ {\rm T} }^{J_{ {\rm T} }} }$ and $\ket{\Psi_{ {\rm p} }^{J_{ {\rm p} }} }$ denote the target and projectile states with their respective angular momenta ${ { J }_{ {\rm T} } }$ and ${ { J }_{ {\rm p} } }$. The total angular momentum ${  J }$ is obtained by coupling target and projectile angular momenta ${ { J }_{ {\rm T} } }$ and ${ { J }_{ {\rm p} } }$. In the calculation of $^{19}$F$(p,\gamma)$$^{20}$Ne, we assume $^{12}$C as the inert core. The core potential is given by the Woods-Saxon potential. For the two-body interactions, the Furutani-Horiuchi-Tamagaki finite-range two-body force is used~\cite{Furutani78,Furutani79}. More details of the one-body core potential and the two-body interaction between valence particles are given in the Supplemental Material~\cite{supplement}.

As the model spaces for targets and the reaction composites providing cross sections are different, they will be separately presented in the two following paragraphs.

For the $^{19}$F target wave functions in GSM, the model space for the protons consists of $psd$ and $f_{7/2}$ partial waves.
Since the low-lying states of $^{19}$F are well bound, the $p_{1/2}$,  $d_{3/2}$ and $d_{5/2}$ partial waves are represented by six harmonic oscillator (HO) states to discretize each $\ell_j$ scattering continuum. The oscillator length of 1.75 fm is adopted for the HO states. For the $p_{3/2}$ and $s_{1/2}$ partial waves, five HO states are used for each $\ell_j$.
One HO state is taken for the $f_{7/2}$ partial wave. For neutrons, $0p_{1/2}$, $0d_{5/2}$, and $1s_{3/2}$ single-particle (s.p.) states are included. We recall that the $0s_{1/2}$, $0p_{3/2}$ shells do not belong to the model space, which are fully occupied and form the $^{12}$C core. Thus, the GSM calculations in the Slater determinant (SD) representation are performed in 29 shells for protons: 18 shells $p_{1/2}$, $d_{3/2}$ and $d_{5/2}$, ten shells $p_{3/2}$, $s_{1/2}$ shells, one $f_{7/2}$ shell, and in three shells for neutrons: $p_{1/2}$, $d_{5/2}$, and $s_{1/2}$. Furthermore, the SD basis is truncated by allowing at most two nucleons to occupy the scattering states.

The GSM-CC channel states in $^{20}$Ne are then built by the coupling of the low-lying GSM states of $^{19}$F with the proton in partial waves: $s_{1/2}$, $p_{1/2}$, $p_{3/2}$, $d_{3/2}$, $d_{5/2}$ and $f_{7/2}$. Three lowest 1/2$^{+}$, 5/2$^{+}$, 9/2$^{+}$, 7/2$^{-}$, 9/2$^{-}$ states, two 7/2$^{+}$, and one 3/2$^{+}$, 1/2$^{-}$, 3/2$^{-}$, 5/2$^{-}$ states of the target are taken. For the calculation of the capture reaction, the bound, resonant and scattering states have to be involved, which are incorporated through the Berggren ensemble generated by the Woods Saxon basis potential. Its diffuseness, radius, central and spin-orbit strength are 0.65 fm, 3.5 fm, 50 MeV (55 MeV for $d$ and $f$ shells), and 7.5 MeV,  respectively. For the proton partial waves, $0p_{1/2}$, $0d_{5/2}$, $1s_{3/2}$, $0d_{3/2}$, $0f_{7/2}$ resonant s.p. state, and 60
non-resonant s.p. continuum states in each Berggren contour $\mathcal{L}^+_{p_{1/2}}$, $\mathcal{L}^+_{p_{3/2}}$, $\mathcal{L}^+_{d_{5/2}}$, $\mathcal{L}^+_{d_{3/2}}$,  and $\mathcal{L}^+_{f_{7/2}}$ are used. For the first four contours, each contour contains three segments connecting the points: $k_{\text{min}}$=0.0, $k_{\text{peak}}=0.15-i0.01$ fm$^{-1}$, $k_{\text{middle}}=0.3-i0.01$ fm$^{-1}$ and $k_{\text{max}}$=2.0 fm$^{-1}$. The contour $\mathcal{L}^+_{f_{7/2}}$ consists of three segments: $k_{\text{min}}$=0.0, $k_{\text{peak}}=0.475-i0.015$ fm$^{-1}$, $k_{\text{middle}}=0.9-i0.015$ fm$^{-1}$ and $k_{\text{max}}$=2.0 fm$^{-1}$, so as to incorporate its resonant pole located at $k=0.474-i0.011$ fm$^{-1}$. Thus, the total number of channel states for $^{20}$Ne is 9, 22, 30, 26, 37, and 46 for $0^+$, $1^+$, $2^+$, $3^+$, $1^-$, $2^-$, and $3^-$ states respectively. The numerical convergence and uncertainties of the GSM-CC calculation are discussed in details in the Supplemental Material~\cite{supplement}.

\begin{figure}[htb]
\includegraphics[width=0.9\linewidth]{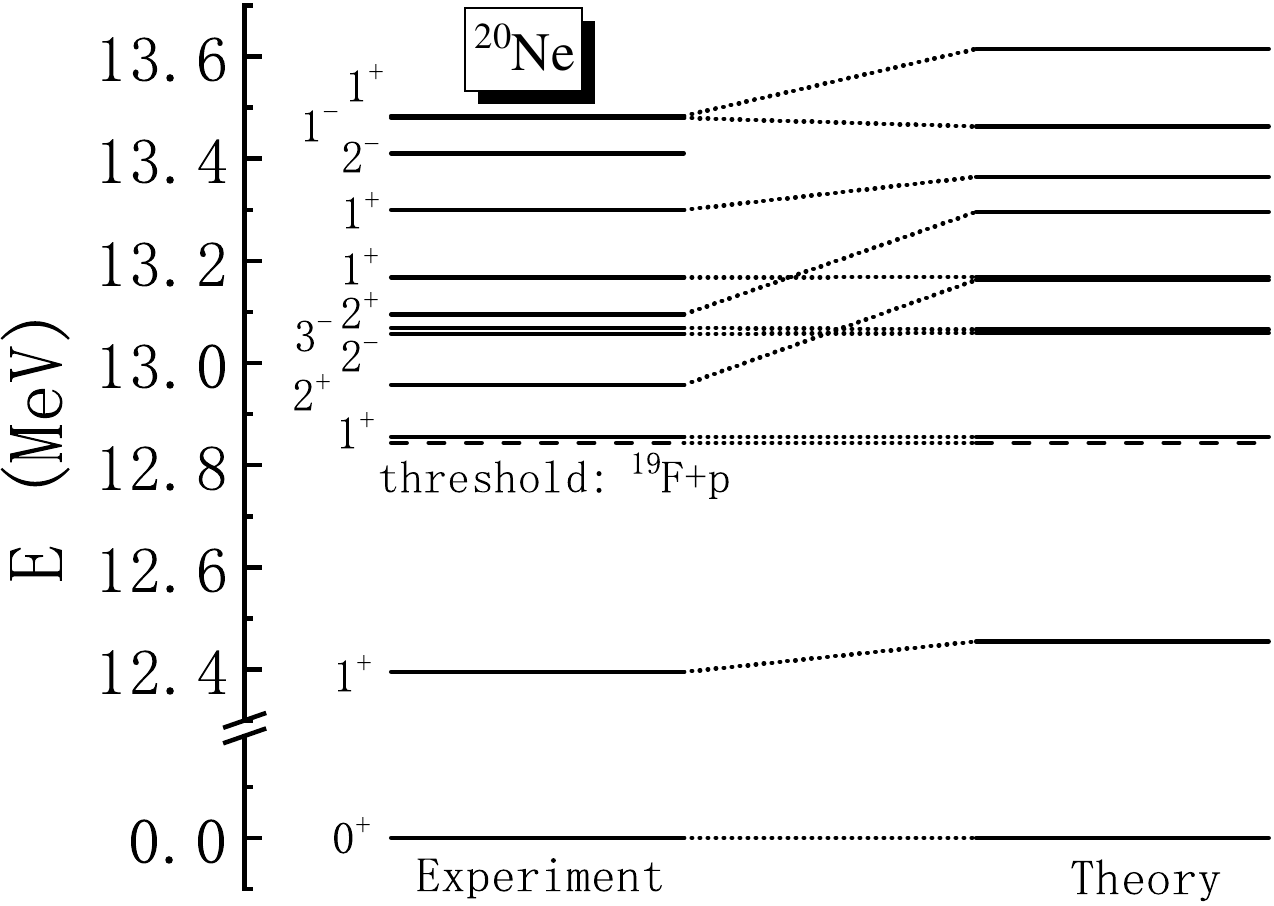}
\caption{The GSM-CC energy levels of $^{20}$Ne are compared with the experimental data~\cite{nndc,TILLEY1998249,Kious90,CO08,deBoer,JUNA}. Most of the energy levels are contained in the NuDat 3.0 database~\cite{nndc}. Remaining experimental data are from recent experiments (see a discussion in the Supplemental Material~\cite{supplement}.  The energy of the ground state is set to zero, and the energies of excited states are shown relative to the ground state. }
\label{fig-1b}
\end{figure}

\textit{Results --} The GSM-CC and experimental spectra of $^{20}$Ne are shown in Fig.~\ref{fig-1b} for states around the proton emission threshold, which are important for the $^{19}$F$(p,\gamma)$$^{20}$Ne reaction.
For $1^+$ states, there are two resonances in the vicinity of the proton emission threshold: the near-threshold resonance at 11 keV~\cite{Kious90} and the resonance at 323 keV above the threshold~\cite{nndc}. Experimental energies of these states are well reproduced by the GSM-CC calculation. There are also higher lying $1^+$ resonances known at 457 keV and 641 keV above the threshold~\cite{nndc}, for which the GSM-CC calculation yields slightly higher energy. In the vicinity of the proton emission threshold one finds also $2^-$ state at 212.7(10) keV and $3^-$ state at 225.2(10) keV~\cite{JUNA}, which are well reproduced by GSM-CC. The calculated resonance energy of $1^-$ state is also very close to the experimental value $E(1^-)_{\exp}$ = 637 keV~\cite{nndc}. Calculated energies of the $2^+$ resonances are slightly above the experimental values but these two states are mainly involved in the $\alpha$ decay of $^{20}$Ne~\cite{nndc} and do not affect the radiative proton capture cross-section~\cite{JUNA}.

The calculated binding energy of $^{19}$F with respect to the core of $^{12}$C is -55.710 MeV, close to the experimental value -55.640 MeV. However, in order for radiative capture reaction cross sections to be precisely calculated in GSM-CC, we introduce the experimental ground-state energy of $^{19}$F in the coupled-channel equations instead of its calculated value. This ensures that the experimental proton separation energy in $^{20}$Ne is exactly taken into account in the cross section calculation.

As the test of calculated wave functions, we compare in Table~\ref{m-moment} experimental and calculated magnetic moments for low lying states in $^{19}$F and $^{20}$Ne. One can see that the GSM calculation reproduce the experimental data quite well.

\begin{table}[ht!]
\caption{ The calculated magnetic moments of low-lying states of $^{19}$F and $^{20}$Ne are compared with the experimental data from the compilation~\cite{stone2014table}. The unit is the nuclear magneton $\mu_\textsc{N} $. }
\label{m-moment}
\begin{ruledtabular}
\begin{tabular}{lcc}
  & Experiment~\cite{stone2014table} &GSM\\
    \hline
  $^{19}$F, 1/2$^+_1$  &+2.628868(8)   &2.823  \\
  $^{19}$F, 5/2$^+_1$  &+3.607(8);3.595(13)   &3.365  \\
  $^{19}$F, 5/2$^-_1$  &0.67(11)   & 0.972 \\
  $^{20}$Ne, 2$^+_1$  & 1.08(8) & 1.094\\
\end{tabular}
\end{ruledtabular}
\end{table}

\begin{figure}[htbp]
	\includegraphics[width=1.0\linewidth]{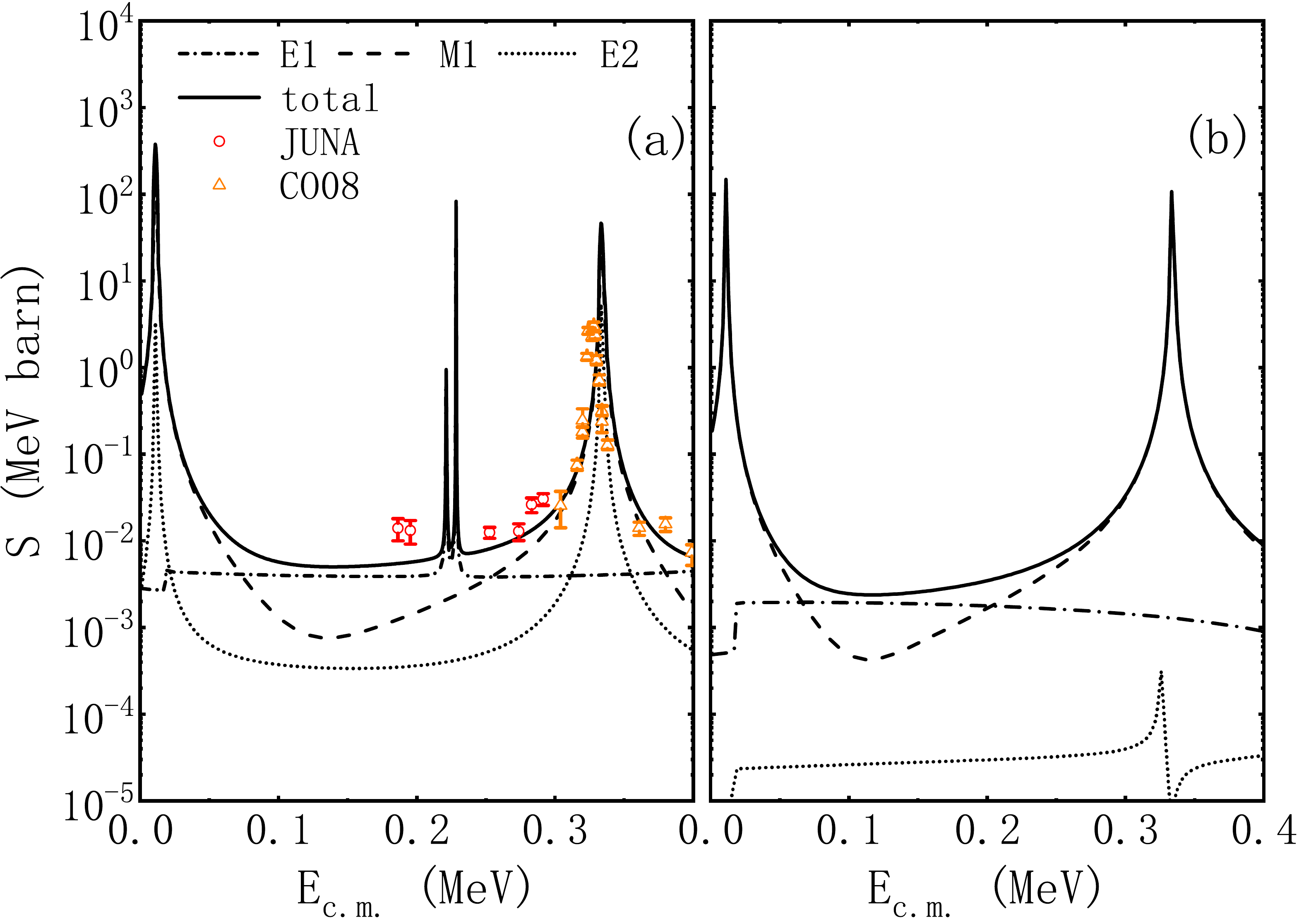}
	\caption{The astrophysical factor of the $^{19}$F$(p,\gamma)$$^{20}$Ne reaction is plotted as a function of proton projectile energy in the $p$ + $^{19}$F center of mass frame.
The lines represent results of the fully antisymmetrized GSM-CC calculation. The left panel is for the radiative capture to the first excited state of $^{20}$Ne, and right panel is for the capture to the ground state. The experimental data from Refs.~\cite{JUNA} and \cite{CO08} are labeled as ``JUNA'' and ``CO08'', respectively.}
	\label{fig-2}
\end{figure}

The proton radiative capture cross section to the final state of $^{20}$Ne with total angular momentum ${ {J}_{f} }$ is obtained by integrating the differential cross-sections.
For calculating the antisymmetrized matrix element of electromagnetic operators, we use the same method as explained in Ref.~\cite{Fossez15}. Short-range and long-range parts in electromagnetic operators are separated~\cite{Fossez15,dong17,GSMbook}. As the short-range operators are only important in the nuclear zone, they are handled by the HO expansion. Long-range operators extend in the asymptotic region. They are treated by the complex rotation for the infinite range of the radial integrals~\cite{Fossez15,dong17,GSMbook}. As a result, the infinite-range of electromagnetic operator has been fully considered by the radial integrals of GSM-CC matrix elements.

Electromagnetic transitions connect the scattering states of $^{20}$Ne with the capturing state in $^{20}$Ne. For the direct capture to the ground state of $^{20}$Ne, referred to as $(p,\gamma_0)$, we consider the incoming states of $p$ + $^{19}$F with spin $J^{\pi}_{\rm in} = {1}^-$ for the $E1$ transition, $J^{\pi}_{\rm in} = {1}^+$ for the $M1$ transition, and $J^{\pi}_{\rm in} = {2}^+$ for the $E2$ transition. For the capture to the first excited state of $^{20}$Ne, referred to as $(p,\gamma_1)$, we take into account the incoming states with $J^{\pi}_{\rm in} = {1}^-,~2^-,~3^-$ for the $E1$ transition,  $J^{\pi}_{\rm in} = {1}^+,~2^+,~3^+$ for the $M1$ transition, and $J^{\pi}_{\rm in} =0^+,~{1}^+,~2^+,~3^+$ for the $E2$ transition. To compute the cross section for radiative capture reaction very precisely, we use the experimental ground state energy of $^{19}$F in GSM-CC instead of the one by GSM, so that the experimental proton separation energy in $^{20}$Ne is exactly taken into account in the cross section calculation.

For the radiative capture cross section of the charged particle, it is convenient to use the astrophysical $S$ factor so that the effect of the Coulomb interaction is removed.
For $E1$ and $E2$ transition calculations, the effective charges originating from the recoil correction of the center of mass~\cite{Hornyak75,YKHo88} are used. No effective charge is needed for $M1$ transitions. In Fig.~\ref{fig-2}, the astrophysical $S$ factors for the proton radiative capture on $^{19}$F to the first excited state (panel (a)) and the ground state (panel (b)) of $^{20}$Ne are shown The separate contributions from $E1$, $M1$ and $E2$ transitions, are also shown.

\begin{figure}[htbp]
	\includegraphics[width=1.0\linewidth]{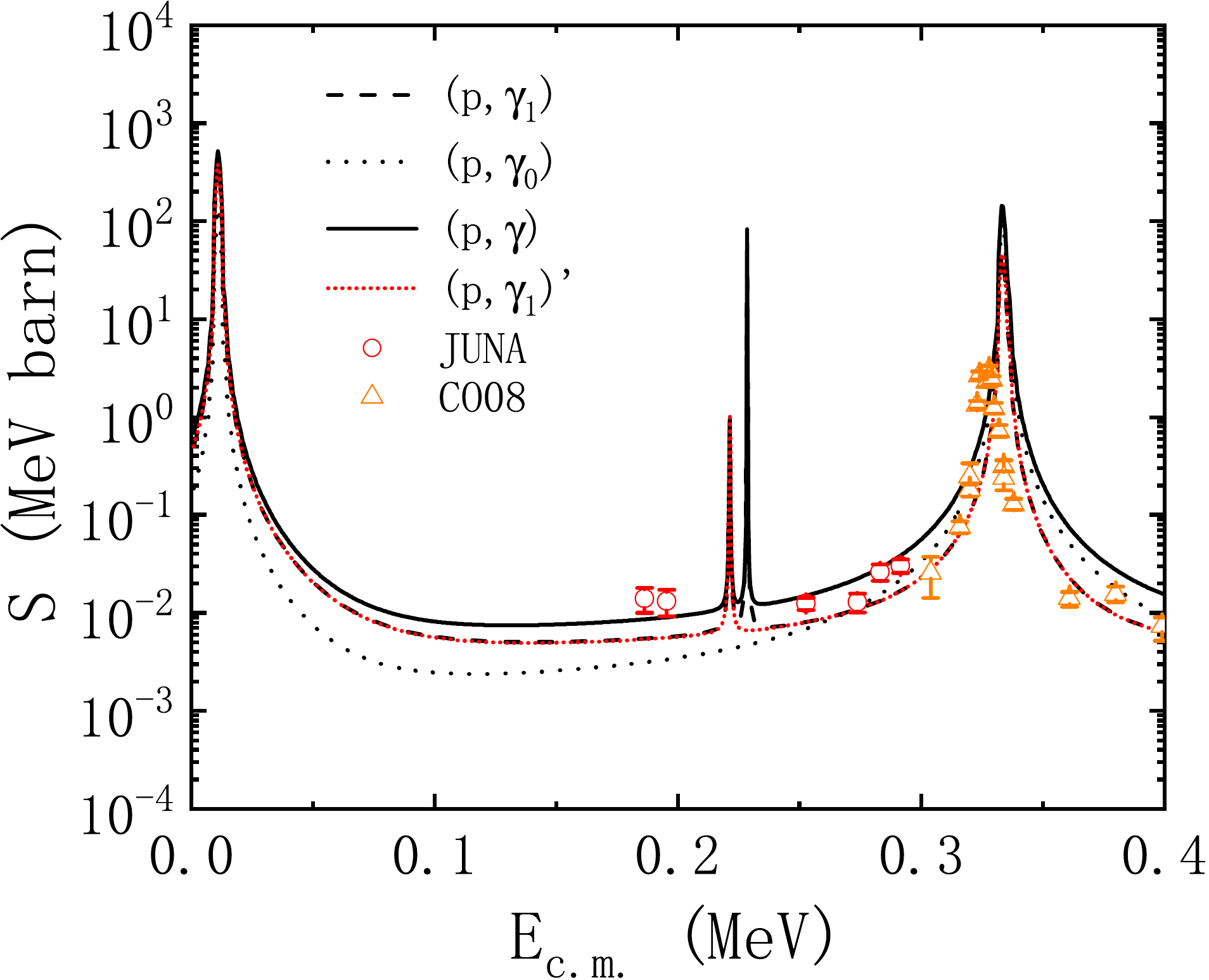}
	\caption{The same as Fig.~\ref{fig-2}, but for the calculated astrophysical factor of the proton capture reaction to the first excited state (dashed line), ground state of $^{20}$Ne (dotted line), and the summation of them (solid line), respectively. The short dotted red line is for capturing to the first excited state without taking into account the $f_{7/2}$ proton partial wave. }
	\label{fig-3}
\end{figure}

The total $S$ factor, i.e., the sum of the $S$ factor for capturing to the ground and the first excited state, is shown in Fig.~\ref{fig-3}. The existing experimental data~\cite{JUNA,CO08} for the capture to the first excited state of $^{20}$Ne are given for the comparison.

JUNA has found a large strength for $3^-$ resonance at $E_{\rm c.m.}=$ 225.2 keV~\cite{JUNA}. In our calculation, a high peak of $3^-$ resonance is also produced. The $3^-$ resonance in GSM-CC is dominated by the channel states with the coupling to the $f_{7/2}$ proton partial wave. For the test, we give the S factors without the $f_{7/2}$ proton partial wave, in Fig.~\ref{fig-3}. One can see that the S factors without $f_{7/2}$ partial wave are barely changed except for the missing peak from the $3^-$ resonance. As seen in Fig.~\ref{fig-2} (a), there is a peak caused by $2^-$ resonance, roughly two orders of magnitude smaller than that of the nearby $3^-$ resonance. According to the measurements~\cite{CO08,JUNA}, the strength of this $2^-$ resonance is less than 10\% of that of the $3^-$ resonance.

From the spectra of $^{20}$Ne shown in Fig.~\ref{fig-1b}, one can see that there is one near-threshold and one resonant $1^+$ states below $E_{\rm c.m.}\approx$ 350 keV. It is extremely difficult to obtain the experimental data at very low $E_{\rm c.m.}$.
At $E_{\rm c.m.}\approx$ 11 keV, a very high peak of the near-threshold $1^+$ state is seen in GSM-CC. The peak caused by the $1^+$ resonance at $E_{\rm c.m.}\sim$ 323 keV is also obtained. The data points for $(p,\gamma_1)$ given by JUNA Collaboration, are in the non-resonant region with $E_{\rm c.m.}\approx$ 180--300 keV~\cite{JUNA}  (see Fig.~\ref{fig-2} (a)). In this region, the calculated $S$ factors are slightly smaller than the JUNA data~\cite{JUNA}. In the energy range around $E_{\rm c.m.}\sim$ 323 keV, the calculated $S$ factors are slightly higher than the experimental data given in Ref.~\cite{CO08}.

\begin{figure}[htb]
\vskip 0.5truecm
\includegraphics[width=0.9\linewidth]{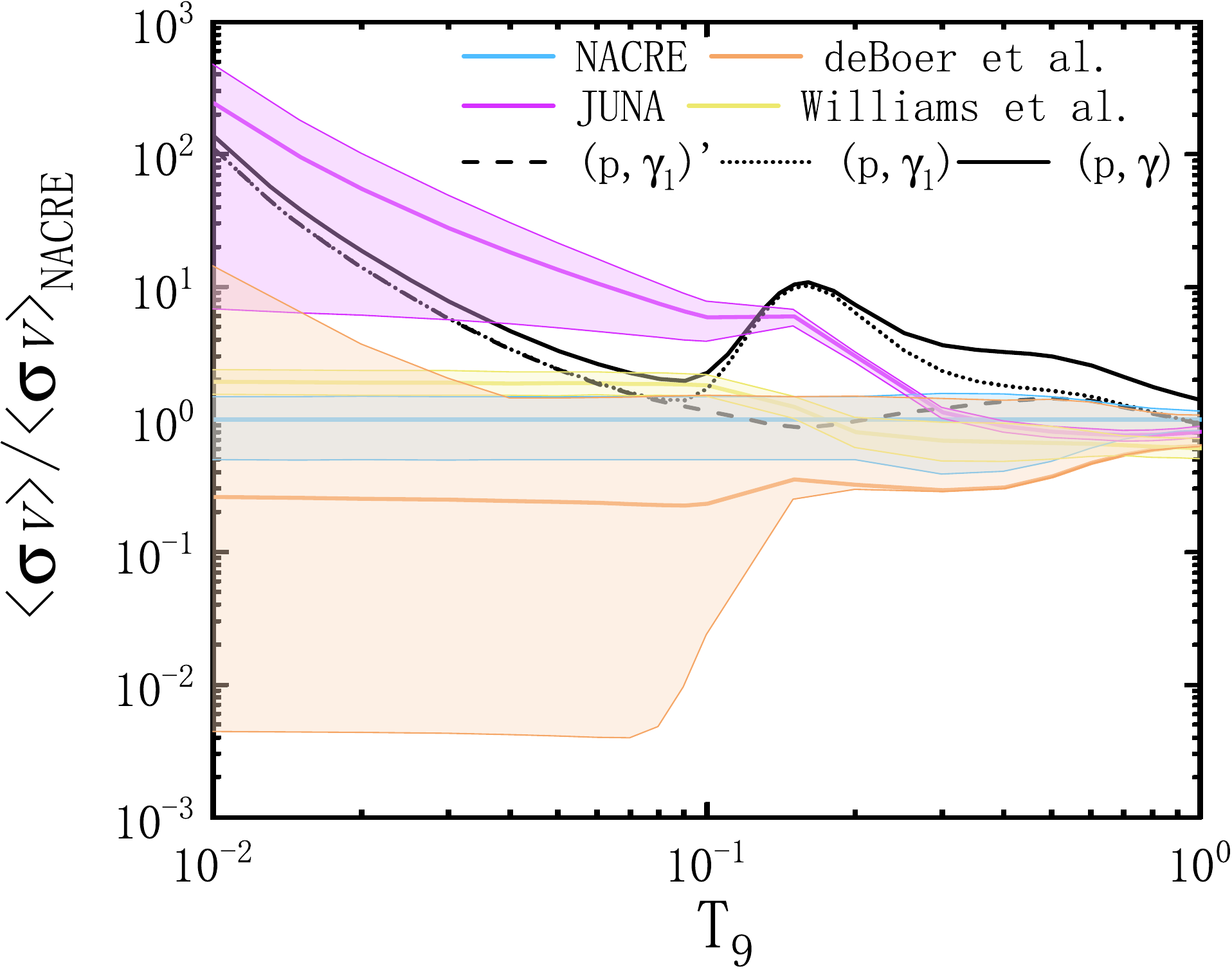}
\caption{ The ratio of the $^{19}$F$(p,\gamma)$$^{20}$Ne reaction rate relative to the NACRE~\cite{nacre} rate as a function of the temperature $T_9$ calculated by GSM-CC. The solid black line is for the calculated total rate, with the capture to the ground and the first excited states of $^{20}$Ne. The dotted black line is the calculated rate for the first excited state only. The dashed black line is for the capture to the first excited state without the inclusion of $f_{7/2}$ proton partial wave. The corresponding ratio of the reaction rate from JUNA~\cite{JUNA}, deBoer \textit{et al.}~\cite{deBoer}, Williams  \textit{et al.}~\cite{Williams} are also shown for comparison.}
\label{fig-5}
\end{figure}

In Fig.~\ref{fig-2} (b) for the capture to the ground state  $(p,\gamma_0)$, there are mainly two peaks, caused by the near-threshold and the resonant $1^+$ states through the $M1$ transition. There is also a very small peak from $2^+$ resonance, several orders of magnitude lower than the $1^+$ resonance, which is negligible in the total S factor. In the JUNA measurement~\cite{JUNA}, it has been found that below $E_{\rm c.m.}\approx$ 322 keV, the strength of the capture to the first excited state $(p,\gamma_1)$ is stronger than that of the capture to the ground state $(p,\gamma_0)$. In the calculation of GSM-CC, $(p,\gamma_1)$ component for $E_{\rm c.m.}<$ 300 keV has also a larger $S$ factor than $(p,\gamma_0)$.

The reaction rate of $^{19}$F$(p,\gamma)$$^{20}$Ne  at around 0.1 GK is of extreme importance for the origin of calcium made from pop III stars, and for the validation of stellar evolution models~\cite{Keller14}.
Thus, in the following we will focus on the study of thermonuclear reaction rate for the temperature ranging from 0.01--1 GK.

Figure~\ref{fig-5} shows the ratio between the GSM-CC reaction rate and the NACRE recommended reaction rate~\cite{nacre}. For comparisons, the ratio of reaction rates obtained in  earlier studies of different groups are also shown. In Fig.~\ref{fig-5}, we give the GSM-CC results obtained by using the total $S$ factors and the $S$ factors corresponding to $(p,\gamma_1)$ only. For a comparison, the ratio of reaction rates obtained by neglecting the $f_{7/2}$ partial wave in GSM-CC is shown as well. We may notice that the inclusion of $f_{7/2}$ is of critical importance for temperatures in the interval from 0.1 GK to $\sim$0.5 GK.

Using the NACRE recommended rate for $^{19}$F$(p,\gamma)$$^{20}$Ne and $^{19}$F$(p,\alpha)$$^{20}$Ne reactions, the predicted calcium abundance would be 2 orders of magnitude lower than the observation in the most metal-poor stars, which causes the calcium abundance puzzle. In the recent analysis by $R$-matrix method, the derived rate becomes smaller, making the calcium abundance problem even worse~\cite{deBoer}. The rate by Williams \textit{et al.}~\cite{Williams} are slightly larger than the NACRE, since their measurement is also sensitive to the proton capture to the ground state while the previous experiments are only sensitive to the transition to the first excited state. However, in the latest JUNA data, the reaction rate of $^{19}$F$(p,\gamma)$$^{20}$Ne is enhanced by a factor of 5.4--7.4 for temperatures around 0.1 GK, which is close to explaining the origin of calcium in the early stars.
\begin{figure}[htb]
\vskip 0.5truecm
\includegraphics[width=0.9\linewidth]{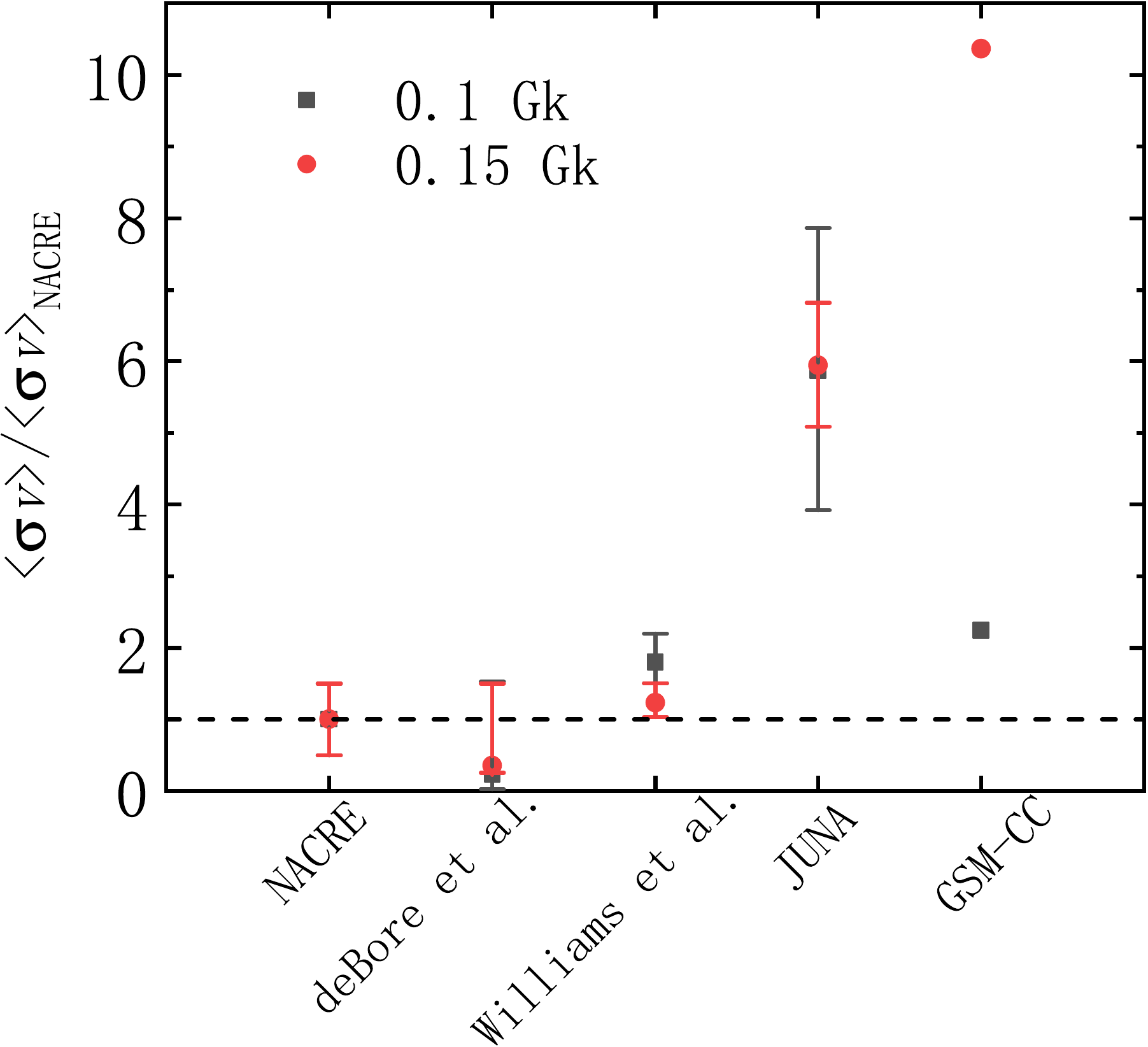}
\caption{ The ratio of the $^{19}$F$(p,\gamma)$$^{20}$Ne rate relative to the NACRE~\cite{nacre} rate at temperature of 0.1 and 0.15 GK.}
\label{fig-6}
\end{figure}
At $T_9$ = 0.01, the calculated rate by GSM-CC is over 100 times larger than the NACRE recommendation and is similar to the rate reported by JUNA. At the temperature ranging from 0.06 GK to 0.08 GK, the rate in the present work is lower than that from JUNA, but is still larger than the rate by NACRE. This is mainly because the calculated $S$ factors for the non-resonant region at low $E_{\rm c.m.}$ are smaller than the experimental data of the JUNA Collaboration. At 0.1 GK, the rate by GSM-CC is enhanced by a factor of 2.24 relative to the NACRE recommendation (see Fig.\ref{fig-6}).

By comparing the curves using the $S$ factors of $(p,\gamma_1)$  with or without the $f_{7/2}$ partial wave, one can see that the increase of the reaction rate at around 0.1 GK is mainly caused by the $3^-$ resonance. The reaction rate calculated by GSM-CC reaches the JUNA value at 0.12 GK, and then overshoots significantly the reaction rate given by JUNA at higher temperature. At around 0.15 GK it is about 1.7 times larger than the JUNA rate (see Fig. \ref{fig-6}). The main reason of the higher rate is caused by the $2^-$ resonance at $E_{\rm c.m.}=$ 213 keV, which is missing in the $R$-matrix calculation by JUNA~\cite{JUNA}. At 1 GK, the discrepancies among the reaction rates become quite small. The rate by JUNA is smaller than both the NACRE rate and the GSM-CC rate.

\textit{Conclusions --} In this work, we have studied the $^{19}$F$(p,\gamma)$$^{20}$Ne reaction using the GSM-CC approach. In general, the agreement between the calculated $S$ factors and the previous data is quite good. Based on the $S$ factors calculated in GSM-CC, the astrophysical reaction rate of $^{19}$F$(p,\gamma)$$^{20}$Ne is derived. The GSM-CC rate is over two times higher than NACRE recommendation at 0.1 GK, and is lower than the rate by JUNA below 0.1 GK. The calculated rate approaches JUNA rate at 0.12 GK, and becomes significantly larger than the JUNA rate at higher temperatures. Overall, our rates are close to those obtained by JUNA Collaboration, and higher than previous evaluations. Thus, based on the present GSM-CC studies, $^{19}$F$(p,\gamma)$$^{20}$Ne reaction cross section is sufficiently large to overcome the $^{19}$F$(p,\alpha)$$^{16}$O reaction which is transforming $^{19}$F back into the CNO cycle. Hence,  $^{19}$F$(p,\gamma)$$^{20}$Ne breakout reaction from the CNO cycle may possibly explain the calcium abundance in the first generation stars.

\textit{Acknowledgements --} We wish to thank Michael Wiescher for useful comments. This work has been supported by the National Natural Science Foundation of China under Grant Nos. U2067205, 12275081, 12175281, 11605054, and 12347106. G.X.D. and X.B.W. contributed equally to this work.

\end{document}